\documentclass[sigconf]{acmart}
\copyrightyear{2025}
\acmYear{2025}
\setcopyright{rightsretained}
\acmConference[CIKM '25]{Proceedings of the 34th ACM International Conference on Information and Knowledge Management}{November 10--14, 2025}{Seoul, Republic of Korea}
\acmBooktitle{Proceedings of the 34th ACM International Conference on Information and Knowledge Management (CIKM '25), November 10--14, 2025, Seoul, Republic of Korea}\acmDOI{10.1145/3746252.3760913}
\acmISBN{979-8-4007-2040-6/2025/11}
\usepackage{graphicx} % Required for inserting images
\usepackage{xcolor}
\usepackage{listings}
\usepackage{caption}
\usepackage{hyperref}

\lstdefinestyle{custom}{
  basicstyle=\ttfamily\small,
  backgroundcolor=\color{white},
  keywordstyle=\color{blue},
  commentstyle=\color{gray},
  showstringspaces=false,
  breaklines=true,
  moredelim=**[is][\colorbox{yellow}]{@}{@}
}


\title{Powering Job Search at Scale: LLM-Enhanced Query Understanding in Job Matching Systems}

\author{Ping Liu}
\affiliation{
 \institution{LinkedIn Corporation}
 \city{Mountain View}
 \state{CA}
 \country{USA}
}
\email{piliu@linkedin.com}

\author{Jianqiang Shen}
\affiliation{
 \institution{LinkedIn Corporation}
 \city{Mountain View}
 \state{CA}
 \country{USA}}
\email{jshen@linkedin.com}

\author{Qianqi Shen}
\affiliation{
 \institution{LinkedIn Corporation}
 \city{Mountain View}
 \state{CA}
 \country{USA}}
\email{qishen@linkedin.com}

\author{Chunnan Yao}
\affiliation{
 \institution{LinkedIn Corporation}
 \city{Mountain View}
 \state{CA}
 \country{USA}}
\email{chyao@linkedin.com}

\author{Kevin Kao}
\affiliation{
 \institution{LinkedIn Corporation}
 \city{Mountain View}
 \state{CA}
 \country{USA}}
\email{kkao@linkedin.com}

\author{Dan Xu}
\affiliation{
 \institution{LinkedIn Corporation}
 \city{Mountain View}
 \state{CA}
 \country{USA}}
\email{dnxu@linkedin.com}

\author{Rajat Arora}
\affiliation{
 \institution{LinkedIn Corporation}
 \city{Mountain View}
 \state{CA}
 \country{USA}}
\email{rajarora@linkedin.com}

\author{Baofen Zheng}
\affiliation{
 \institution{LinkedIn Corporation}
 \city{Mountain View}
 \state{CA}
 \country{USA}}
\email{bzheng@linkedin.com}

\author{Caleb Johnson}
\affiliation{
 \institution{LinkedIn Corporation}
 \city{Mountain View}
 \state{CA}
 \country{USA}}
\email{cajohnson@linkedin.com}

\author{Liangjie Hong}
\affiliation{
 \institution{LinkedIn Corporation}
 \city{Mountain View}
 \state{CA}
 \country{USA}}
\email{liahong@linkedin.com}

\author{Jingwei Wu}
\affiliation{
 \institution{LinkedIn Corporation}
 \city{Mountain View}
 \state{CA}
 \country{USA}}
\email{jingwu@linkedin.com}

\author{Wenjing Zhang}
\affiliation{
 \institution{LinkedIn Corporation}
 \city{Mountain View}
 \state{CA}
 \country{USA}}
\email{wzhang@linkedin.com}

\begin{document}

\begin{abstract}
Query understanding is essential in modern relevance systems, where user queries are often short, ambiguous, and highly context-dependent. Traditional approaches often rely on multiple task-specific Named Entity Recognition models to extract structured facets as seen in job search applications. However, this fragmented architecture is brittle, expensive to maintain, and slow to adapt to evolving taxonomies and language patterns. In this paper, we introduce a unified query understanding framework powered by a Large Language Model (LLM), designed to address these limitations. Our approach jointly models the user query and contextual signals such as profile attributes to generate structured interpretations that drive more accurate and personalized recommendations. The framework improves relevance quality in online A/B testing while significantly reducing system complexity and operational overhead. The results demonstrate that our solution provides a scalable and adaptable foundation for query understanding in dynamic web applications.
\end{abstract}



\maketitle

\section{Introduction}

Query understanding \cite{chang2020query} is the process of interpreting the underlying intent behind a user's search query in search engines and recommender systems. In most search engine scenarios, queries are short, free-form text strings -- often just a few tokens. One core objective is to extract salient information, map it to structured facets or attributes \cite{guo2009named, wang2024enhancing}, and apply relevant filters to generate personalized, context-aware, and relevant results.

Modern recommender systems face significant challenges in query understanding, especially in domains like job search, where users are seeking life-changing opportunities \cite{lu2013recommender, LinkRetrieval2024, liu2025scalable, liu2025linksage}. While we use job recommendations as the running example, these challenges generalize across many domains.
First, user intent is often implicit rather than explicitly stated. Though signals from the user's profile, activity history, and preferences can help infer intent, effectively integrating these diverse sources remains a complex modeling task.
Second, queries can be ambiguous without sufficient context. For instance, ``find me a job in Naples'' could refer to Naples, Florida or Naples, Italy. Disambiguating such queries requires robust geographic and contextual reasoning.
Third, queries are frequently incomplete or noisy, leading to inaccurate candidate matches and a suboptimal user experience. Overcoming these issues are critical for delivering high-quality, intent-aligned recommendations.

% \begin{table}[tb]
% {
%     \caption{Challenges of query understanding in job search recommender system}
%     \begin{tabular}{l|r}
%     \hline
%         Category &  Example Queries \\ \hline 
%         New Taxonomy   & \textit{LLM Tutor} \\
%         Self-reference & \textit{Find the retail jobs in my area} \\ 
%         Typo           & \textit{Health xare jobs} \\ 
%     \hline
%     \end{tabular}
%     \label{tab.case_study}
% }    
% \end{table}

% what are we doing in this paper?
% In our current system, the query understanding stack relies on separate Named Entity Recognition (NER) models \cite{li2020survey} to identify key facets such as company names, job titles, geographic locations, and etc. However, this approach has notable limitations: maintaining multiple specialized models is costly, and the system struggles to adapt to the dynamic evolution of language and terminology. To overcome these limitations, we propose an LLM-powered query understanding framework tailored to the job search domain. This unified solution leverages the reasoning and generalization capabilities of a large language model to enhance facet extraction, intent disambiguation, and the overall job search candidate relevance. Our preliminary online A/B tests show that this framework significantly improves matching accuracy in candidate selection, reduces serving latency, and simplifies system maintenance by replacing several components with a single production-ready LLM. Specifically, our contribution in this paper is summarized as follows: 
As one of the largest job platforms globally, LinkedIn has traditionally relied on separate Named Entity Recognition (NER) models \cite{li2020survey} to extract key facets such as company names, job titles, and geographic locations. While effective for narrow tasks, this architecture is costly to maintain and struggles to keep up with the evolving nature of language and user intent. To overcome these limitations, we present a unified, LLM-powered query understanding framework. By leveraging the reasoning and generalization capabilities of LLMs, our approach enables end-to-end facet extraction, intent disambiguation, and semantic refinement within a single model. Preliminary online A/B testing demonstrates improvement in candidate matching accuracy, reduced serving latency, and simplified system complexity -- achieved by consolidating multiple legacy components into one streamlined LLM solution. Specifically, our contribution in this paper is summarized as follows: 

    \textit{\textbf{LLM-Based Query Understanding Framework}}: We present a unified, production-ready framework that refines free-text search queries using a single LLM, jointly modeling intent interpretation and facet extraction for improved personalization and relevance. 

    \textit{\textbf{Practical Deployment Strategy and Guide}}: We offer actionable guidance for real-world LLM deployment, including infrastructure design, latency optimization, and cost-efficiency strategies, particularly valuable to teams operating under resource constraints.

    \textit{\textbf{Demonstrated Business Impact}}: We report strong performance improvements in offline evaluations and online A/B tests in the large-scale job matching platform of LinkedIn, showing gains in relevance quality and system efficiency.

The remainder of the paper is organized as follows: Section \ref{sec.related_work} reviews related work. Section \ref{sec.framework} presents our framework, followed by model selection and training details in Section \ref{sec.model}. Experimental results are discussed in Section \ref{sec.experiment}, and we summarize the conclusion in Section \ref{sec.conclusion}.
% Structure in this paper

% https://lucid.app/lucidchart/516b7f3f-8678-4cfa-b20e-fed91f4112f0/edit?viewport_loc=288%2C189%2C2004%2C998%2C0_0&invitationId=inv_6f5adffe-0464-414b-9450-8dff6e602da6 
\begin{figure*}[tb]
    \centering
    \includegraphics[width=0.83\textwidth]{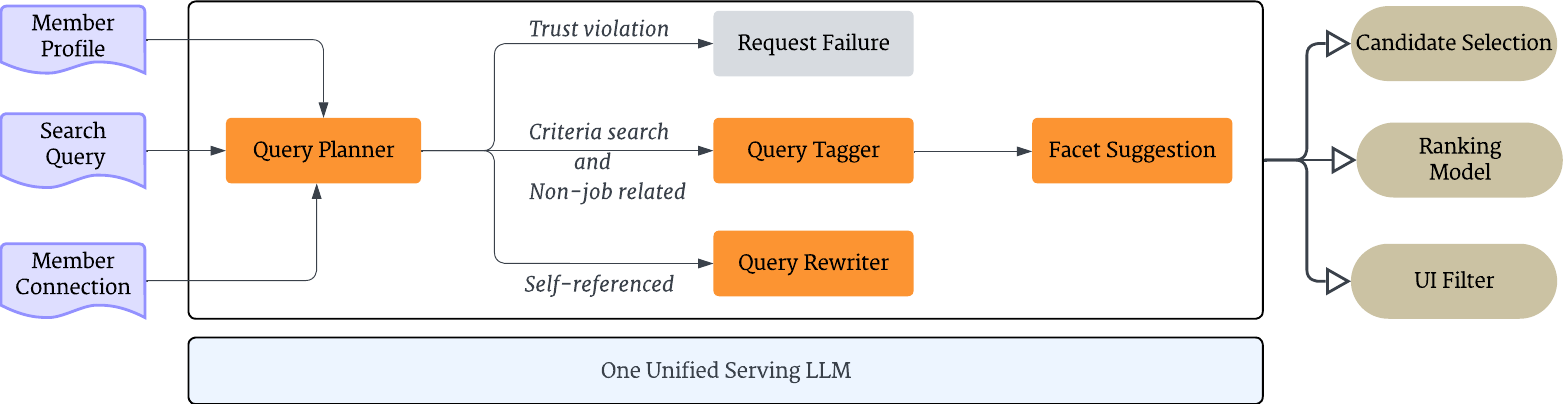}
    \vspace{-4pt}
    \caption{The workflow of our query understanding framework. The components in the center are powered by a single unified LLM. The outputs generated by the query understanding module are consumed by downstream applications.}
    \vspace{-3pt}
    \label{fig.workflow}
    \Description[The pipeline of framework.]{The pipeline of framework.}
\end{figure*}

\section{Related Work}
\label{sec.related_work}

\textbf{Query Understanding in Search Engine.} 
Query understanding aims to interpret user queries for more relevant recommendations. Recent studies show that large language models (LLMs) can significantly enhance this process in recommender systems \cite{dai2024enhancing, abe2025llm, luo2024exploring, zhang2020query, srinivasan2022quill, luo2022query}. JD \cite{dai2024enhancing} proposes a three-stage approach, including domain-specific pre-training, supervised fine-tuning, and reinforcement learning with recall-based rewards to achieve notable gains in recall and conversion. Amazon \cite{luo2024exploring} applies multi-task learning to incorporate query understanding into product ranking. Our framework extends these ideas with a more unified and robust design capable of handling a wider range of production scenarios.
% QUILL \cite{srinivasan2022quill} boosts LLM performance with retrieval-augmented context such as titles and URLs.

\textbf{Tool Calling and Multi-Agent Systems.}
Recent development has positioned LLMs as agents in search and recommendation systems, enabling advanced tool usage and reasoning beyond traditional pipelines \cite{zhang2025survey}. Their knowledge and generalization capabilities have shown promise in addressing cold-start and data sparsity challenges \cite{shu2024rah}. For example, RecMind \cite{wang2024recmind} employs an LLM-based agent with external knowledge and planning tools, outperforming prior zero- and few-shot baselines on benchmark tasks. Similarly, MACRec \cite{wang2024macrec} introduces a multi-agent framework that tackles diverse recommendation challenges through agent collaboration. Despite these advances, applications of such agent-based frameworks at industrial scale remain limited, especially in billion-user recommendation systems. This work addresses that gap.

\section{Proposed System Framework}
\label{sec.framework}
In this section, we present an overview of the framework designed to enhance  query understanding in the job marketplace. It contains 4 main components, each of which plays a critical role in interpreting user intent and improving the user experience. The general framework is shown in Figure \ref{fig.workflow}, powered by one unified LLM.

\subsection{Query Planner}
The Query Planner \cite{wang2024recmind, zhang2025survey} serves as the central intelligence of the entire framework. It determines how to best route a user's job search query through the recommender system by analyzing intent and mapping it to predefined categories, then directing it to the appropriate downstream components:

\textit{\textbf{Criteria search}}: 
Most search queries fall into this category, characterized by specific conditions, filters, or requirements relevant to the items in the repository. These queries often include attributes such as job title,  location, seniority, company name, and other facets. Such criteria-based searches are routed to the query tagging component, which identifies and annotates key elements of the query to support more accurate candidate matching and ranking.

\textit{\textbf{Self-reference search}}: 
If the query needs a user's unspecified explicit personal information, such as geographic location, work experience, past roles, education, skills etc to successfully perform the search. For example, ``jobs that match my profile'', ``PM jobs that fit my qualifications''. These self-reference queries require the necessary information from the user's profile. 
The planner calls the query rewriter to enrich the query with the necessary information for downstream processing.

\textit{\textbf{Non-job related search}}: 
Some queries may not pertain to job search, and distinguishing them from genuine job-seeking intent can be challenging. To handle this, the query planner flags potentially irrelevant queries but still forwards them to candidate selection and ranking for processing.
For example, a query like ``I want to be a mermaid'' can still lead to relevant job listings such as Underwater Aquatics Performer or Pearl Diver at SeaWorld. This capability demonstrates the system's flexibility and surpasses the limitations of traditional keyword-based job search engines.

\textit{\textbf{Trust violated search}}:
We define several contextual scenarios that constitute trust violations, including the use of offensive language, violent content, discriminatory remarks, self-harm-related queries involving feedback, and other harmful or inappropriate behaviors. When the system detects such contents, the query is immediately denied, and a warning message is displayed to the user: ``This search query may violate our Professional Community Policies. Edit your search to try again''.

\subsection{Query Tagger}
In recommender systems, query tagging \cite{manshadi2009semantic, zuo2023deeptagger} refers to extracting structured attributes from user queries to enable more effective candidate matching. In the job marketplace domain, this task is particularly challenging due to the need for precise, multidimensional matching across attributes such as job title, seniority, industry, and geographic location. Compounding this complexity, the taxonomies for these attributes evolve continuously, limiting the adaptability of traditional methods like Named Entity Recognition (NER), which often fail to handle dynamic, domain-specific terminology.

To address these challenges, we adopt a generative LLM-based approach for facet tagging. The input prompts specify the extraction requirements and enforce a structured output format (e.g., JSON or TypeScript). Our unified LLM model is capable of tagging key job-related attributes such as company, title, and geographic location, as well as additional attributes such as the 'easy apply' feature, application count restrictions, and others that conventional NER or encoder-based models typically lack. The main limitations of NER and encoder-based approaches lie in their limited reasoning capabilities and poor generalization across diverse or complex scenarios. These extracted attributes serve three primary purposes: (a) certain attributes (e.g., location, Easy Apply) are used for attribute-based filtering during the candidate selection stage; (b) textual entities such as title and company are incorporated as input features in both the candidate selection and ranking models.

\subsection{Query Rewriter}
Query rewriter \cite{he2016learning, mo2024aligning} is to modify a user's query to improve the relevance of the matching results. The purpose is to disambiguate the implicit intentions from job search queries and expand synonyms terms to increase the recall of recommended jobs. 
The rewriting logic targets self-referential queries, which often depends on the job seeker's profile, such as education, skills, title, location, industry, or experience. When self-reference is detected, the LLM takes both the query and member profile as input. For instance, it rewrites “Find software engineer jobs near my location” to “Find software engineer jobs near Bay Area, CA” if the member's location is Bay Area. The rewritten query is then passed to candidate selection and ranking models, resulting in more accurate and relevant matches.

\subsection{Facet Suggestion}
Facet suggestion refers to the process of recommending filtering options (facets) that assist users in refining and exploring search results. These facets -- such as industry, company, and job title -- enable users to narrow down results or explore relevant opportunities. The system may also provide personalized facet recommendations based on user intent, search queries, and historical behaviors.
To mitigate the risk of overly broad suggestions, we constrain the taxonomy for each facet type within the prompt and only recommend a facet if it is possibly related to the query and aligned with the user’s profile. In our preliminary A/B testing, facet suggestion was activated only when an industry was explicitly mentioned in the query. Upon detecting an industry, the LLM suggests related industries, which are then surfaced in the search user interface.

\begin{table*}[tb]
\small
\centering
\caption{Precision and recall of various tools using Pretrained and Fine-tuned LLMs.}
\vspace{-4pt}
\begin{tabular}{lcccccccccc}
\toprule
& \multicolumn{2}{c}{\textbf{Query planner}} 
& \multicolumn{2}{c}{\textbf{easy\_apply\_tool}} 
& \multicolumn{2}{c}{\textbf{date\_posted\_tool}} 
& \multicolumn{2}{c}{\textbf{num\_applicants\_tool}} 
& \multicolumn{2}{c}{\textbf{job\_in\_network\_tool}} \\
\cmidrule(lr){2-3} \cmidrule(lr){4-5} \cmidrule(lr){6-7} \cmidrule(lr){8-9} \cmidrule(lr){10-11}
\textbf{Model} & \textbf{Precision} & \textbf{Recall} & \textbf{Precision} & \textbf{Recall} & \textbf{Precision} & \textbf{Recall} & \textbf{Precision} & \textbf{Recall} & \textbf{Precision} & \textbf{Recall} \\
\midrule
Pretrained LLM & 0.810 & 0.900 & 0.100 & 1.000 & 0.620 & 0.910 & 0.650 & 0.680 & 0.080 & 0.960 \\
Fine-tuned LLM & 0.994 & 1.000 & 1.000 & 1.000 & 0.923 & 1.000 & 1.000 & 1.000 & 1.000 & 1.000 \\
\bottomrule
\end{tabular}
\label{tab:tool_performance_llms}
\vspace{-2pt}
\end{table*}

\section{Model Training and Serving}
\label{sec.model}

In this section, we present the details of our modeling approach, including the multi-task prompts strategy for the LLM fine-tuning and the rationale behind our infrastructure choices.

\subsection{Modeling}

As established in Section~\ref{sec.framework}, we propose a unified LLM architecture to serve all generative tasks in our production environment. This approach consolidates distinct functionalities into a single, efficient model. The LLM is optimized for two primary generation modalities: (1) Agentic Tool Calling, and (2) Text Rewriting. This unified strategy is designed to minimize deployment overhead and maintain consistency across user-facing features.

For our fine-tuning paradigm, we adopt a multi-task instruction tuning approach, inspired by the T0 framework \cite{sanh2021multitask}. 
% An illustrative example of a multi-task prompt is provided in Figure~\ref{fig.prompt}. 
A critical challenge in multi-task learning is mitigating catastrophic forgetting and managing interference between disparate task gradients \cite{luo2023empirical}. To address this, we first curated our task datasets by strategically upsampling to ensure balanced class distribution and task representation. To accelerate data collection, we also followed the methodology outlined in \cite{taori2023stanford} and supplemented with synthetic samples generated by prompting GPT to realistically modify existing queries for tool testing, conditioned on member information. All samples were reviewed by human annotators for quality assurance.
We then investigated two data scheduling strategies: (1) \textbf{Heterogeneous Batching}: In this baseline approach, each mini-batch is constructed by randomly sampling from the aggregated dataset, mixing instances from all tasks; (2) \textbf{Homogeneous Batching}: Our proposed approach constructs each mini-batch exclusively from a single task. The task for each batch is selected randomly or based on a predefined curriculum.

Our empirical results demonstrated that Homogeneous Batching outperformed Heteregeneous Batching. We hypothesize this is because disjoint batches provide a more stable gradient signal for each task-specific optimization step, preventing conflicting updates within a single batch.  This approach achieved performance on individual tasks that was comparable to dedicated, single-task models.
Due to strict production constraints requiring a P95 end-to-end inference latency under 600ms, our backbone model selection was heavily influenced by performance–efficiency trade-offs. Smaller models (<2B parameters) met latency targets and achieved high recall, but exhibited lower precision and frequent hallucinations in structured outputs, especially for tool-calling. Larger models (>7B) offered higher fidelity but were computationally impractical.

We choose \href{https://huggingface.co/Qwen/Qwen2.5-1.5B-Instruct}{Qwen2.5-1.5B model} \cite{bai2023qwen} as the backbone to fine-tune the tasks described in Section~\ref{sec.framework}. %An example prompt is illustrated in Figure~\ref{fig.prompt}.  
We compared two model tuning methodologies: Supervised Fine-Tuning (SFT) \cite{howard2018universal} that uses full-parameter updates to maximize flexibility in learning structured outputs, and DPO \cite{rafailov2023direct}, a reinforcement learning algorithm that optimizes on chosen/rejected response pairs.
Both achieved similar performance, with DPO slightly reducing hallucinations.
We ultimately chose SFT for its superior control over structured outputs -- crucial for agentic tasks requiring strict adherence to JSON schema. DPO’s preference-based objective occasionally compromised format fidelity. Additionally, SFT enabled a simpler MLOps pipeline \cite{anil2022factory} and faster iteration cycles, making it better suited for production deployment. The training objective of SFT is to minimize the cross-entropy loss, between the predictions and the ground-truth responses, defined as:

\vspace{-2pt}
\begin{equation}
\mathcal{L}_{\text{SFT}} = -\sum_{i=1}^N \log P_\theta(y_i \mid x_i),
\end{equation}

where $x_i$ denotes the input prompt for the $i$-th training instance, $y_i$ is the corresponding target output (e.g., human-generated response), $P_\theta(y_i \mid x_i)$ is the probability of $y_i$ given $x_i$ under model parameters $\theta$, and $N$ is the total number of training examples.

\subsection{Infrastructure for Online Inference}

The unified LLM is deployed via vLLM \cite{kwon2023efficient} and exposed as a gRPC service. vLLM is selected for its high throughput and low latency, especially with small decoder models sharing long prompt prefixes, making it well-suited for large-scale generative workloads. In-house models are served behind a custom gRPC proxy compatible with the OpenAI Chat Completion API. The decoupled, modular architecture enhances observability, fault isolation, and flexibility. This  evolution from earlier tightly-coupled systems reduces maintenance burdens by abstracting internal logic and allowing for seamless integration of complex request types within the \href{https://www.langchain.com/}{LangChain} framework.

To optimize latency for the global user base of LinkedIn, which operates under stringent low-latency demands of 600ms P95 at thousands of queries per second, streaming mode is enabled for model serving. A custom JSON parser incrementally processes streamed output, allowing for the parallel execution of recognized tools as they are identified. This concurrent execution, combined with vLLM's efficient server-side and client-side batching, significantly improves throughput and reduces end-to-end latency, making the system scalable and responsive.

\section{Experimental Results}
\label{sec.experiment}

% In this section, we focus on offline metrics, online A/B testing results, as well as infra and production observation.

% \begin{figure}[tb]
%     \centering
%     \includegraphics[width=0.45\textwidth]{figures/latency.pdf}
%     \vspace{-5pt}
%     \caption{The comparison of latency benchmark between vLLM and SGLang.}
%     \label{fig.star_graph}
%     \Description[The comparison of latency benchmark between vLLM and SGLang.]{The comparison of latency benchmark between vLLM and SGLang.}
%     \vspace{-12pt}
% \end{figure}

\subsection{Offline Metrics}

In our system, the existing stack only supports limited tagging, such as geographic location and company, therefore, there is limited evaluation against the baseline. We prepare a training and evaluation dataset around $3K \sim 5K$ for each task and label them with dedicated human annotators. Qwen2.5-1.5B-Instruct \cite{bai2023qwen} is fine-tuned with Cosine scheduler \cite{loshchilov2016sgdr} and learning rate $2e^{-5}$. Two H100s with 80GB RAM are used for training with batch size 4 on each device. The average time per training step is around 10 seconds.

\begin{table}[tb]
\small
\centering
\caption{Precision and Recall performance for geographic location and company tools.}
\vspace{-4pt}
\begin{tabular}{lcccc}
\toprule
\textbf{} & \multicolumn{2}{c}{\textbf{location\_tool}} & \multicolumn{2}{c}{\textbf{company\_tool}} \\
\cmidrule(lr){2-3} \cmidrule(lr){4-5}
\textbf{Model} & \textbf{Precision} & \textbf{Recall} & \textbf{Precision} & \textbf{Recall} \\
\midrule
Legacy NER        & 0.934 & 0.894 & 0.688 & 0.710 \\
Fine-tuned LLM       & 0.954 & 0.981 & 0.800 & 0.910 \\
\bottomrule
\end{tabular}
\vspace{-2pt}
\label{tab:location_company_perf}
\end{table}

We compare the precision and recall of geographic location and company in the Table \ref{tab:location_company_perf}. Our fine-tuned lightweight LLM outperforms the legacy models in both precision and recall. We further compare our fine-tuned model against pre-trained models on additional tasks. As shown in Table~\ref{tab:tool_performance_llms}, the fine-tuned model consistently outperforms the pre-trained baseline across all metrics. The performance gap is primarily due to the tendency of smaller pre-trained LLMs to hallucinate in response to examples provided in the prompt, which results in relatively lower precision.

\subsection{Online A/B test}
We conducted a four-week online A/B test to evaluate our new framework. The results show consistent improvements in relevance across multiple metrics. Specifically, the Normalized Discounted Cumulative Gain (NDCG) \cite{wang2013theoretical} increased by 33\% ($p<.05$), while the proportion of poor matches (results deemed irrelevant by product policy) among the top 10 ranked results dropped by 59\% ($p<.05$).
Our system processes roughly 20 queries per second per A100 GPU, with a median latency of 400ms and P95 latency of 600ms. We leverage Multi-Instance GPU to maximize hardware utilization by partitioning each GPU into isolated instances.

By consolidating multiple legacy models into one framework, we reduce maintenance overhead by over 75\% and streamline deployment and versioning.
These improvements are critical for accelerating iteration velocity and supporting scalable product development.

\section{Conclusion}
\label{sec.conclusion}
In this paper, we propose a novel query understanding framework aimed for a billion-scale job search recommender system. 
Our contributions go beyond simply addressing the limitations of traditional search engines. We share practical insights and lessons learned from deploying large-scale systems in a real-world, industrial setting. These include strategies for building scalable infrastructure, optimizing model performance, and maintaining system reliability. Experiments show that our approach delivers significant gains across both offline evaluation metrics and online A/B testing, highlighting its effectiveness and production readiness.

\clearpage
\balance

% Required by ACM and CIKM this year. Otherwise, desk rejected...
\section*{GenAI Usage Disclosure}
GenAI is used for label generation for SFT training. No other GenAI tools were used in any stage of the research, nor in the writing.

\bibliographystyle{ACM-Reference-Format}
\bibliography{main}

\end{document}